\documentclass[%
 reprint,
superscriptaddress,
 amsmath,amssymb,
 aps,
 prx,
]{revtex4-2}
\usepackage{graphicx}
\usepackage{ulem}
\usepackage{dcolumn}
\usepackage{bm}
\usepackage{hyperref}
\usepackage{physics}
\usepackage{orcidlink}
\usepackage{amsfonts}

\newcommand{\indep}{\perp\!\!\!\perp}

\usepackage{xcolor}
\newcommand{\vcq}{ University of Vienna, Faculty of Physics, Vienna Center for Quantum Science and Technology (VCQ) {\&} Research platform TURIS, Boltzmanngasse 5, 1090 Vienna, Austria}
\newcommand{\cdl}{ Christian Doppler Laboratory for Photonic Quantum Computer, Faculty of Physics,  University of Vienna, 1090 Vienna, Austria}

\begin{document}

\title{Towards an Experimental Device-Independent Verification of Indefinite Causal Order}


\author{Carla M. D. Richter\orcidlink{0009-0009-6576-2501}$^{1,*, \dagger}$, Michael Antesberger\orcidlink{0000-0003-0978-4692}$^{1,*}$, Huan Cao \orcidlink{0000-0001-7171-4148}$^{1}$, Philip Walther\orcidlink{0000-0002-4964-817X}$^{1,2}$, Lee A. Rozema \orcidlink{0000-0003-4580-7871}$^{1,\ddagger}$\\
\small\ $^{1}$\textit{\vcq} \\
\vspace{0.6mm}
\small $^{2}$\textit{\cdl} \\
\vspace{1.5mm}
\small $^{*}$\textit{ These two authors contributed equally to this work.} \\
\small \textit{Correspondence to:}\\
\small $^{\dagger}$\textit{ carla.richter@univie.ac.at},
\small $^{\ddagger}$\textit{ lee.rozema@univie.ac.at}\\
}

%
%
%
%

\date{\today}

\begin{abstract}
In classical physics, events follow a definite causal order: the past influences the future, but not the reverse. Quantum theory, however, permits superpositions of causal orders---so-called indefinite causal orders---which can provide operational advantages over classical scenarios. Verifying such phenomena has sparked significant interest, much like earlier efforts devoted to refuting local realism and confirming quantum entanglement.
To date, demonstrations of indefinite causal order have all been based a process called the quantum switch and have relied on device-dependent or semi-device-independent protocols. 
Achieving a device-independent verification of indefinite causal order would imply that nature allows for correlations that do not respect causality, independent of any experimental assumptions or underlying theoretical description of the experiment. To this end, a recent theoretical development introduced a Bell-like inequality that allows for fully device-independent verification of indefinite causal order in a quantum switch.
Here we implement this verification by experimentally violating this inequality. In particular, we measure a value of $1.8328  \pm  0.0045$, which is 18 standard deviations above the Definite Causal Order Bound of $1.75$.
Our work presents the first implementation of a device-independent protocol to verify indefinite causal order, albeit in the presence of experimental loopholes.
This represents an important step towards the device-independent verification of an indefinite causal order, and provides a context in which to identify loopholes specifically related to the verification of indefinite causal order.
\end{abstract}
\maketitle

\section{Introduction}
In a classical understanding of causality, events have a well-defined order in time, meaning that events in the past can only influence those in the future.
Any process with a well-defined causal order will satisfy so-called {causal inequalities}, which impose constraints on temporal correlations generated by causality-respecting processes \cite{oreshkov2012quantum}.
Quantum mechanics appears to allow for events to occur in a superposition of orders, such processes are said to have an an indefinite causal order (ICO) \cite{oreshkov2012quantum,Chiribella2013quantum,baumeler2014maximal,baumeler2016space,wechs2021quantum}, which is required to violate a causal inequality. 
Causal inequalities are device independent (DI), meaning that their violation would prove, independent of quantum theory, that nature allows for correlations that do not respect our classical notion of causality.
However, not all processes with an ICO can violate a causal inequality \cite{Branciard2016simplestCausalInequalities}.

For example, the quantum switch \cite{chiribella2022indefiniteTimeDirection}, which can be experimentally implemented \cite{rozema2024experimental,Procopio2015Experimental,rubino2017ExperimentalVerification,Rubino2021Communication,Rubino2022experimentalEntanglement,guo2020experimental,cao2022quantumSimulation,cao2022Semideviceindependent,zhu2023prl,Min23,wei2019experimentalCommunication,goswami2018Indefinite,goswami2020IncreasingCommunication,yin2023experimental,Antesberger2023tomography,guo2025surpassing}, does not violate a causal inequality \cite{Araujo2017purificationPostulate,Purves2021CannotViolate}.
Nevertheless, its ICO has been experimentally confirmed in different device dependent ways, such as demonstrating advantages over causally ordered process \cite{Procopio2015Experimental}, using causal witnesses \cite{rubino2017ExperimentalVerification}, and even performing full process tomography \cite{Antesberger2023tomography}.
Moreover, the quantum switch may also be interesting for applications as it has been shown that it can outperform causally-ordered processes at a wide variety of tasks such as channel discrimination \cite{bavaresco2022UnitaryChannelDiscrimination}, promise problems \cite{Araujo2014ComputationalAdvantage}, communication complexity \cite{Guerin2016communicationComplexity}, noise mitigation \cite{Ebler2018EnhancedCommunication}, various thermodynamic applications \cite{Felce2021IBMswitch,Guha2020Thermodynamic,Simonov2022WorkExtraction}, quantum metrology \cite{Frey2019depolarizingChannelIdentification}, quantum key distribution \cite{SpencerWood2023qkd},  entanglement generation \cite{Koudia2023EntanglementGeneration} and distillation \cite{Dey2023EntanglementDistribution}, among others.
Thus, both for foundational interest and to put the many proposed applications on a solid footing, an unambiguous confirmation that ICO is a physically real phenomenon is essential.
In other words, we wish to treat the quantum process as an untrusted adversary, and perform a DI test to definitively rule out the possibility that the observed correlations result from a classical causal order.

All current demonstrations of ICO in the quantum switch have been device dependent or semi-device independent \cite{Rubino2022experimentalEntanglement, cao2022Semideviceindependent}.
Using such an experiment to claim ICO, is akin to claiming a violation of local realism using device-dependent techniques such as quantum state tomography or entanglement witnesses: this is valid only if all assumptions hold, but it is open to loopholes that could void the experimental conclusions.
In the case of entanglement, this led to a decades-long push to realise a loophole-free violation of local realism via a Bell inequality \cite{Shalm2015strong,Giustina2015Significant,Hensen2015Loophole}.
In the context of ICO, violating a causal inequality is a DI technique which would take the place of a Bell inequality.
When a process has an ICO, one event can generate correlations with other events which occur before and after it.
In all device-dependent experiments carried out so far, assumptions are essentially made about when and where the events occur.
Since it is difficult to ascribe an event to a photon (which always exists in a superposition of different times) traversing an optic, there is a certain amount of ambiguity about what can be claimed with respect to ICO in existing experimental demonstrations.
The device-dependent assumptions in current experiments are made explicit in recent discussions regarding the validity of quantum switch experiments
\cite{Maclean2017mixturesOfCausalRelations, ormrod2022sectorialConstraints,vilasini2024embedding,paunkovic2020distinguishing,vilasini2025events}, showing the relevance of a loophole-free device-independent demonstration of ICO.

Since the the quantum switch is the only ICO process to be experimentally studied but it cannot violate causal inequalities, we would like to perform a different DI experiment to prove that there is no hidden variable description in which the causal order of the quantum switch is well-defined.
To this end, we present an experimental violation of an inequality, introduced by van der Lugt, Barret and Chiribella (VBC) \cite{van2023device}, which correlates a hidden variable with both a fixed causal order in the quantum switch and a second observable that is then used to violate a Bell inequality.
A successful violation of the Bell inequality thus implies that a hidden variable cannot be assigned to the causal order.  This rules out the possibility that any physical theory, consistent with a causally ordered structure, could explain the observations, even when analysed without assuming a specific theory.
In other words, we leverage DI concepts developed for Bell inequalities to provide a DI certification of ICO using the quantum switch.
Although our experiment does not close the usual Bell loopholes (or other loopholes specific to ICO) it presents a significant experimental step towards a loophole-free confirmation of indefinite causal order. 

\begin{figure}[t] 
 \centering
 \includegraphics[width=\linewidth]{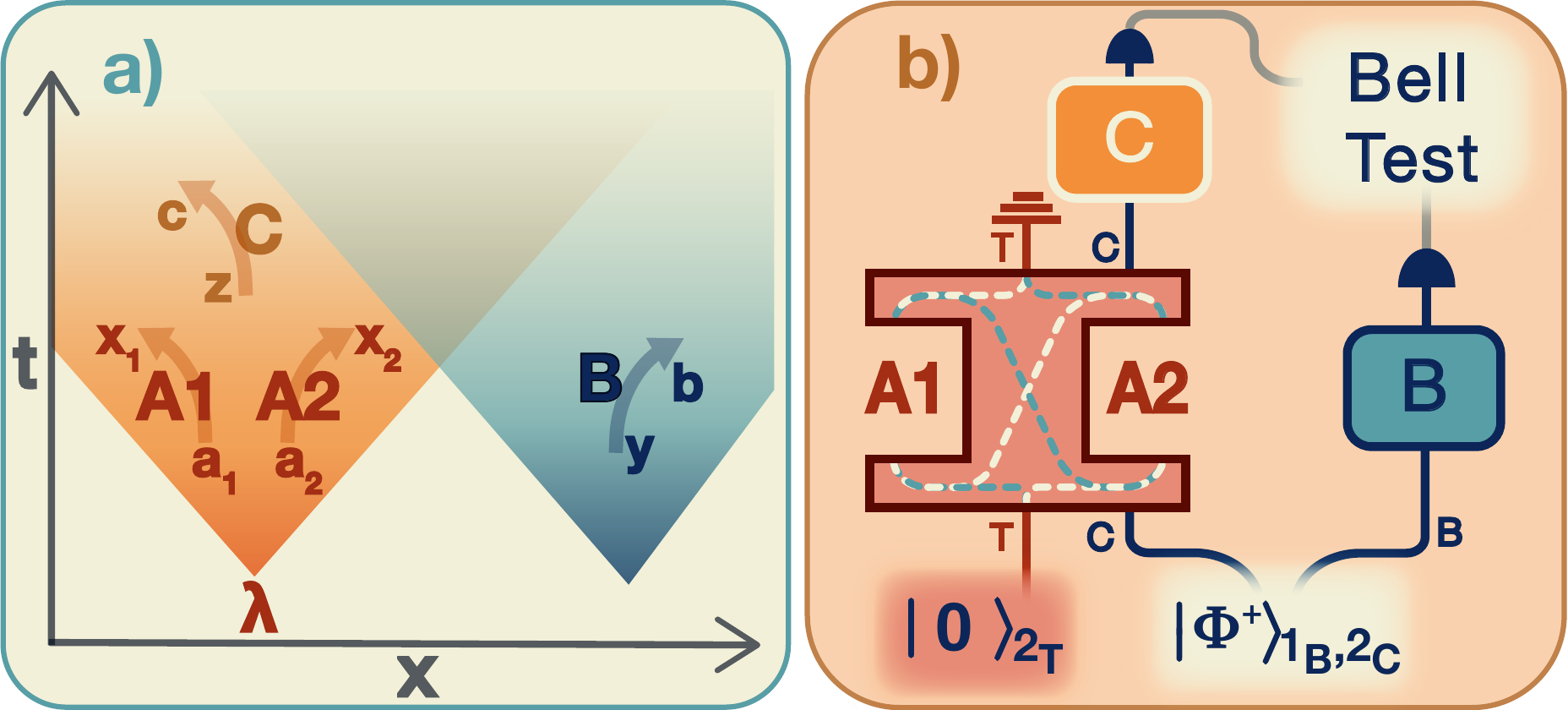}
\caption{\textbf{Causal arrangement for the inequality test.} \textbf{a)}~Spacetime arrangement of the four parties, Alice 1 (A1), Alice 2 (A2), Bob (B), and Charlie (C), who will attempt to violate the inequality. Alice 1 and Alice 2 act before Charlie, while Bob (B) is space-like separated all other participants. 
\textbf{b)} Switch-based protocol to violate the inequality. A Bell state is shared between the control qubit of the quantum switch and Bob. 
Alice 1 and Alice 2 perform measurements on a target qubit in the quantum switch. 
Notice that our notation, which was introduced in \cite{van2023device}, for Alice 1 and 2's settings and outcomes is not the standard Bell test notation.
In particular, $a_{i}$ represents the Alices measurement outcomes as usual. 
However, they always measure in the computational basis. 
Thus no notation is used for their measurement settings. 
Instead, $x_{i}$ denote which state they prepare the post-measurement system in before it leaves their laboratory.
Charlie and Bob's notation is more standard: Charlie measures the control qubit in basis determined by $z$ and receives outcome $c$, and Bob measurement basis is given by $y$ with outcome $b$.}
\label{fig::idea}
\end{figure}

\section{Theory}
The VBC inequality concerns an experiment carried out by four parties: Alice 1 and Alice 2 (who may or may not act in an indefinite causal order), Charlie who acts in the future light cone of the Alices, and Bob, who is space-like separated from the other parties (see. Fig. \ref{fig::idea}a). We then consider DI data generated by these parties.  In particular, on each run of the experiment Alice 1, Alice 2, Bob and Charlie choose their  settings $(x_1, \ x_2,\ y,\ z) \in \{0, 1\}$ and generate outcomes $(a_1, \ a_2,\ b,\ c) \in \{0, 1\}$, respectively. 
The VBC inequality follows from three assumptions: definite causal order, relativistic causality, and free intervention.

The definite causal order assumption introduces a hidden variable $\lambda$, which takes a definite value  on every run of the experiment, with each value of $\lambda$ corresponding to a fixed causal order between the parties. One does not need precisely specify how $\lambda$ acts, rather, we can simply assume that its causal influence exists a priori. 
The relativistic causality assumption states that the causal order, influenced by $\lambda$, must conform to the light-cone structure in which the four parties operate. In particular, since Bob is space-like separated from all other parties, his settings and outcomes are independent of the other three parties, and vice versa.
Moreover, since Alice 1 and Alice 2 act before Charlie, their outcomes and settings are independent of Charlie's actions. 
With this in mind, the hidden variable $\lambda$ can take only two values $\lambda \in \{1, 2\}$. 
For $\lambda = 1$, the order is Alice 1, Alice 2, and then Charlie.
While when $\lambda = 2$, the overall order is Alice 2, Alice 1, and then Charlie. 
Finally, the free-intervention assumption states has two parts. First, it states that all parties can choose their settings freely, independent of variables outside their future light cone, and independent of $\lambda$. Formally this means that the input–output correlations can be represented as:
\begin{equation}
    p(a_1a_2bc \mid  x_1x_2yz) = \sum_{\lambda \in \{1,2\}}p(\lambda)p(a_1a_2bc \mid  x_1x_2yz\lambda).
    \label{eq::ProbDis}
\end{equation}
The second part of the free-intervention assumptions is that, for a given $\lambda$, all the settings are statistically independent of
any outcomes of parties outside their causal order given by $\lambda$.
This can be formally defined by defining the following additional statistical independencies. 
First, we require no signalling between Bob and the other parties as $a_1 a_2 c \indep_p y$ and $b \indep_p x_1 x_2 z$. 
Where the symbol $\indep_p$  defines statistical independence; for example, $a_1 a_2 c \indep_p y$ means $\sum_b p(a_1 a_2 b c\mid  x_1x_2yz) = \sum_b p(a_1 a_2 b c\mid  x_1x_2y'z)$.
Next, as Charlie is time-like separated from Alice 1 and Alice 2, there are two more relevant independencies to define. 
For $\lambda = 1$, Alice 1 is placed in the causal past of Alice 2, which means $a_1 b \indep_p x_2$ and $a_1 a_2 b \indep_p z$. On the other hand, $\lambda = 2$ gives us causal order with the Alices reversed, which requires $a_2 b \indep_p x_1$ and $a_1 a_2 b \indep_p z$.

With this in place, VBC proved that the correlations between the four parties are bounded by:
\begin{equation}
\begin{split}
    p(b=0, a_2 = x_1 \mid y = 0) 
    + p(b=1, a_1 = x_2 \mid y = 0) \\
    +~p(b \oplus c = yz \mid x_1 = x_2 = 0) 
    \leq \frac{7}{4}
\end{split}
\label{eq::LCI}
\end{equation}

We can understand VBC's inequality by considering it to be two separate games: the ``causal order game,'' quantified by the first two terms, and the usual CHSH game, given by the third term.
The causal order game is played whenever Bob chooses the setting $y=0$. The game is won if his outcomes are correlated with the causal order of the Alices.
In particular, the first term quantifies the causal order with $\lambda=1$ in which Alice 1 acts first.
This is achieved by checking if $a_2=x_1$, such that Alice 1 can signal to Alice 2 (Alice 2's outcome $a_2$ is correlated with Alice 1's setting $x_1$) and correlating this with Bob's outcome $b=0$.
The second term is similar, but now it correlates Bob's other measurement outcome $b=1$ to the causal order where Alice 2 acts first and $\lambda=2$.
Note that mathematically, these two terms are bounded by $1$.
Moreover, when they sum to one $p(b=0, a_2 = x_1 \mid y = 0)  + p(b=1, a_1 = x_2 \mid y = 0)=1$, then there are perfect correlations between Bob's measurement outcomes and the hidden variable $\lambda$ defining the causal order.
In this case, Bob and the Alices win the causal order game.

When the parties win the causal order game it means that $\lambda$ is correlated with Bob's measurement outcomes.
We know from Bell's theorem that if $\lambda$ is indeed a classical variable, then Bob's measurement outcomes cannot violate a CHSH inequality with Charlie.
Thus to check the nature of $\lambda$ Bob play the CHSH game with Charlie.

The CHSH game is played whenever the settings of the Alices agree (\textit{i.e.} when they both attempt to send each other $0$).
This is quantified by the third term of Eq. \ref{eq::LCI}, which evaluates at the probability of Bob and Charlie to win the CHSH game.  \textit{i.e.} a referee gives Bob and Charlie the bits $y$ and $z$ they need to generate outputs $b$ and $c$, respectively, that satisfy the following relation $b \oplus c = yz$.
If $\lambda$ is a classical variable Bob and Charlie should be able to win the CHSH game with a probability of at most $\frac{3}{4}$.
So in total the VBC inequality is bounded by $\leq 1+\frac{3}{4}=\frac{7}{4}$.
This bound must hold for any process with a definite causal order. In this case, even if Charlie and Bob share an entangled state and violate the CHSH game in the third term, then the first two terms must decrease to compensate for this.
For Bob to simultaneously win the causal order game with the Alices and the CHSH game with Charlie both entanglement and ICO are required: the causal order of the Alices must be truly indefinite such that they can win the causal order game without destroying the quantum entanglement required to win the CHSH game with Charlie.
This bound can be formalized and generalized for imperfect correlations in a fully DI manner \cite{van2023device}. 

\begin{figure*}[t]
  \centering
  \includegraphics[width=\textwidth]{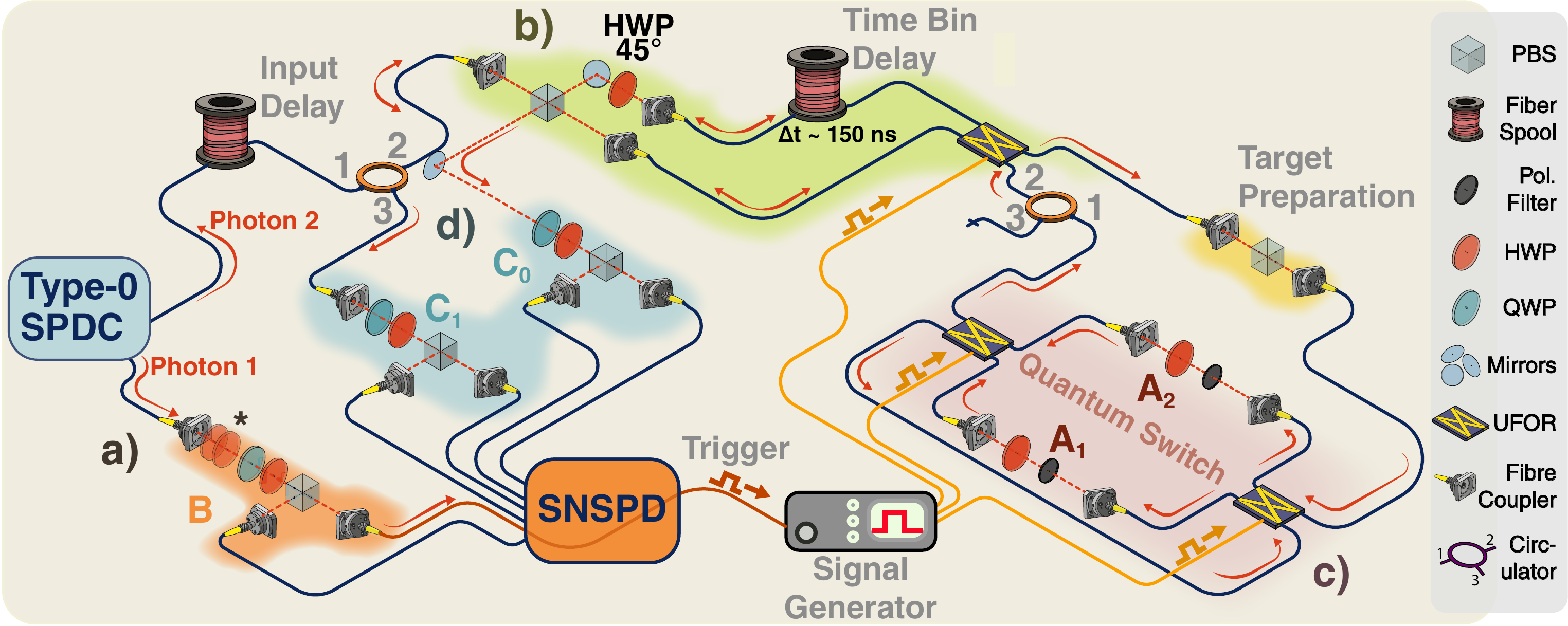}
  \caption{\textbf{Experimental schematic:} A type-0 spontaneous parametric downconversion (SPDC) source generates polarization entangled photon pairs in the $\ket{\phi^+}$ state. \textbf{a)} Photon 1 of the pair is sent to Bob’s measurement stage ($B$), represented by the orange-shaded area. Therein it is measured in a standard polarization measurement stage consisting of a quarter-waveplate (QWP), half-waveplate (HWP) and a polarizing beamsplitter (PBS).  Its detection signal initiates the electrical trigger signal that allows the ultra-fast optical router (UFOR) to switch at the correct time.\textbf{b)} Photon 2 passes an optical circulator before entering the green area, where the polarization qubit is deterministically converted into a time-bin qubit using an imbalanced Mach-Zhender-like interferometer opening on a PBS and closing on an UFOR. For additional. To further refine the polarization, photon 2 passes through a PBS in the target-preparation stage. \textbf{c)} The time-bin qubit then serves as the control for the time-bin quantum switch indicated by the pink shaded region.Here Alice 1 and Alice 2 ($A_1, \ A_2$) perform measurement on the target qubit $a_1, a_2$ and re-prepare the state in the settings $x_1, x_2$ in the polarization DOF using a linear polariser and a HWP. After passing through the quantum switch, the photon travels in the opposite direction, ensuring that both time bins arrive simultaneously at the PBS. \textbf{d)} Depending on their spatial path and polarization, the photons are guided from here to one of Charlie’s measurement stages, $C_0$ or $C_1$ (shaded blue region). Charlie's two polarization measurement devices allow him to implement a complete set of measurements on the control and target qubits. \textbf{*} Two HWPs in $B$ allow the preparation of any Bell state as the input state, enabling additional measurements to test the effect of noise on the VBC inequality.}
    \label{fig::Setup}
\end{figure*}

To see how the quantum switch can be used to violate VBC's inequality, consider the schematic shown in Fig. \ref{fig::idea}b.
The quantum switch, represented as the red shaded areas, takes two input quantum operations ($A_1$ and $A_2$), and applies them to a target system dependent on the state of a control qubit: if the control qubit is in $\ket{0}$ the gates are applied in the order $A_2 A_1$, while they are applied in the order $A_1 A_2$ when the control qubit is in $\ket{1}$.
The control qubit of the quantum switch is entangled in a  $\ket{\Phi^+}$-Bell-state with an ancilla qubit that is sent to Bob.
Alice 1 and Alice 2 are placed inside the quantum switch, and Charlie performs measurements on the control qubit after the switch. The target qubit is prepared in $\ket{0}$ and is discarded after the switch.
Bob and Charlie then play the CHSH game: Bob measures in the $Z$ basis if $y=0$ and in the $X$ basis if $y=1$;
Charlie measures in $X+Z$ basis if $z=0$ and in $X-Z$ if $z=1$.
Inside the switch, both Alices always measure in computational basis to produce outputs $a_i$. They then prepare their outgoing qubit according to their randomly-chosen setting $x_i$: when $x_i=0$ Alice $i$ prepares the state $\ket{0}$ and when $x_i=1$ they prepare  $\ket{1}$.  
In this way they can attempt to signal to each other, and we can check for successful signalling when $a_2=x_1$ or $a_1=x_2$. 
 Note, the notation for Alice 1 and 2 differs from a standard CHSH notation.  In our case, $x_i$ are not the measurement setting but preparation settings; \textit{i.e.} $x_i$ dictates which state Alice tries to send to the other Alice. $a_i$ are measurement outcomes, like usual but the measurements are always performed in the computational basis.
In this configuration, when Bob measures the ancillary qubit in the computational basis, $y=0$,  he will collapse the control qubit of the quantum switch such that if $b = 0$ Alice 1 acts before Alice 2,
whereas if $b = 1$, Alice 2 acts before Alice 1. 
Thus, when $y=0$, with the Alices' measurements and preparations described above, Alice 1 and 2 can signal. In particular, each of the first two terms of Eq.\ref{eq::LCI} will be $\frac{1}{2}$. For the last term, when $x_1=x_2$, the switch effectively acts as an identity channel on the control and target qubits. 
This means that Bob and Charlie will share the input state $\ket{\Phi^+}$ after the switch. 
Thus they can win the CHSH game with a probability given by the Tirelson bound of $\frac{1}{2}+\frac{\sqrt{2}}{4}\approx 0.8536$. 
The net effect is that QM predicts that the setup shown in Fig. \ref{fig::idea}b) will violate the VBC inequality with a value of $1.8536>1.75$.


\section{Experiment}
\begin{figure}[t]
    \centering
    \includegraphics[width=\linewidth]{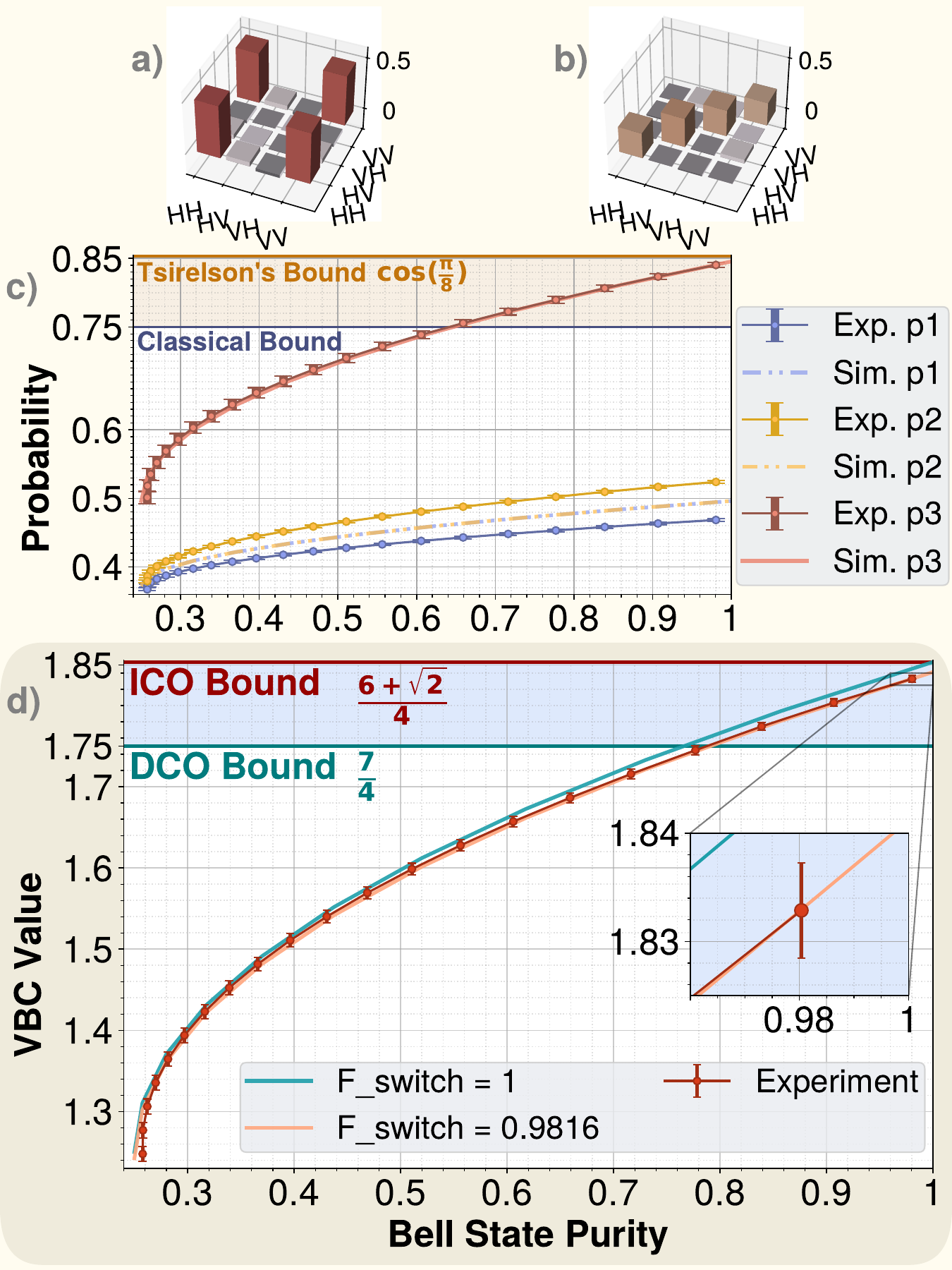}
    \caption{\textbf{Experimental violation of a local causal inequality.}
     \textbf{a)} Density matrix of the two photon input state, as it is generated by our SPDC-Source with a purity of $0.98036 \pm 0.00034$. \textbf{b)} Maximally mixed input state with a purity of $ 0.25777 \pm 0.00015$ created by generating depolarising noise from experimental data via eq.\ref{eq::depol}. \textbf{c)} Effect of depolarising noise, applied on the two photon input state, on the values of the individual term probabilities $p_1, p_2, p_3$.  \textbf{d)} Experimental violation of the VBC inequality,  and simulation of the maximum possible VBC violation as a function of the purity of the input Bell states created by applying different amounts of depolarising noise. Here, the defined causal order (DCO) bound and the indefinite causal order (ICO) bound are marked in color, with the latter representing the maximum achievable value.}
    \label{fig::Data}
\end{figure}

To experimentally violate VBC's inequality, we use a photonic implementation of the quantum switch.
The photonic switch has been built in various forms, using different degrees of freedom (DOFs) of single photons to encode the control and target systems \cite{rozema2024experimental,Procopio2015Experimental,rubino2017ExperimentalVerification,Rubino2021Communication,Rubino2022experimentalEntanglement,guo2020experimental,cao2022quantumSimulation,cao2022Semideviceindependent,zhu2023prl,Min23,wei2019experimentalCommunication,goswami2018Indefinite,goswami2020IncreasingCommunication,yin2023experimental,Antesberger2023tomography,guo2025surpassing}. 
Here, we extend a recent realization of the quantum switch that uses the temporal DOF for the control qubit and the polarization DOF for the target qubit \cite{Antesberger2023tomography}. This configuration is particularly advantageous as it enables the use of simple, waveplate-based gate operations by Alice 1 and Alice 2 in the switch, and the time-bin encoding allows for passive phase stability in Charlie's interferometric measurement of the control qubit.
To generate entanglement between the control qubit and Bob's ancillary qubit we start by generating polarization-entangled photon pairs at telecom wavelength ($\lambda = 1550$ nm) in the $\ket{\Phi^+}$ state using a Type-0 spontaneous parametric down-conversion (SPDC) source in a Sagnac interferometer.
We measure a fidelity to the target state of $0.97197 \pm 0.00066$ (Fig. \ref{fig::Data}a), with a coincidence rate of $\approx 7$ kHz, measured directly from the source.
One photon of the pair is directly sent to Bob (photon 1, Fig. \ref{fig::Setup}a), who will perform arbitrary polarization measurements  $y$ to receive output $b$. The detection signal of photon 1 is then used as a trigger for the synchronization of the 2x2 ultra-fast optical routers (UFOR). Meanwhile the other photon (photon 2) is sent toward the quantum switch. (Note that this does not follow the space-time diagram of Fig. \ref{fig::idea}a, and thus opens a loophole which we will discuss later.)
To create entanglement between Bob's polarization qubit (encoded in photon 1) and the control qubit of the switch (encoded in photon 2), we transfer the polarization DOF of photon 2 to a time-bin DOF. 
We accomplish this by sending photon 2 to an imbalanced Mach-Zehnder--like interferometer setup (Fig. \ref{fig::Setup}b) that opens on a polarizing beam splitter (PBS) and closes on an UFOR \cite{luiz2021fiber}.

If photon 2 is horizontally polarized $\ket{H}$, it is transmitted through the PBS and takes the short path and is then routed into the switch by the UFOR resulting in the early time-bin state $\ket{E}$. When it is vertically polarized $\ket{V}$, on the other hand, it reflects into the long path.
Therein a half-wave plate (HWP) rotates the polarization state from $\ket{V}$ to $\ket{H}$ before the photon experiences a $\tau\approx 150$ ns fibre delay and is routed into the switch in the late time-bin state $\ket{L}$. The HWP ensures both time-bin components share the same polarization and disentangles the polarization of photon 2 from the polarization of photon 1. 
This completes the transfer of entanglement into the time DOF, and sets photon 2's polarization to $\ket{H}$ in both time bins so that it can be used as the target qubit.
More precisely, the state at this point in the experiment is $\ket{\Psi}=\frac{1}{\sqrt{2}}(\ket{HE}_{1_{\text{B}},2_{\text{C}}} + \ket{VL}_{1_{\text{B}},2_{\text{C}}})\otimes\ket{H}_{\text{2}_{\text{T}}}$, where the subscript $1_{\text{B}}$ indicates the polarization encoded ancillary qubit in photon 1, $2_{\text{C}}$ refers to the time-bin encoded control qubit and $2_{\text{T}}$ denotes the polarization-encoded target qubit of photon 2.

Upon entering the quantum switch, the two switch (Fig. \ref{fig::Setup}c) UFORs are synchronized to route the early time bin to Alice 1 and then Alice 2, while the late time bin is routed to Alice 2 and then Alice 1.
A comprehensive description of the quantum switch is provided in in the Supplementary Material 1 and in Ref. \cite{Antesberger2023tomography}. 
Within the quantum switch, Alice 1 and Alice 2, perform projective measurements $a_i$ in the computational basis and subsequently re-prepare the the target system in $\ket{H}$ or $\ket{V}$, depending on their setting $x_i$.
Experimentally, the measurements are achieved by using linear polarisers and checking to see if the photon is transmitted or not.
After the polariser, the state is re-prepared with suitable waveplates dictated by $x_i$.
Note that the final target state is only read out after the switch operation is complete (i.e. we must check if the photon arrives at Charlie's measurement station), which introduces a second loophole by not enforcing the space-time structure of Fig. \ref{fig::idea}a).

After the switch, Charlie measures the time-bin control qubit, which requires him to interfere the early and late time-bin states. We do so using the same Mach-Zhender--like interferometer in reverse; \textit{i.e.} the UFOR now sends the early time bin into the path with the delay and the late time bin into the shorter arm such that the two time bins meet on the polarizing beamsplitter.
Using the same interferometer results in passive interferometric phase stability, see Ref. \cite{Antesberger2023tomography} for more details. 
Since the polarization state is in general changed in the switch, we must consider how the joint state of the control and target qubits is mapped onto a four-dimensional Hilbert space spanned by two path modes and two polarization modes. 
By placing independent polarization tomography modules in each output path of the PBS where the time bins interfere—each consisting of a PBS, QWP, and HWP, with four detectors (one detector at each output PBS output port, Fig. \ref{fig::Setup}d).  Depending on the former time-bin state, that is now encoded in the path, the polarization, and Charlie’s settings $z_i$, the photon ends up in one of the four detectors. This allows a complete set of measurements on both qubits to be performed, yielding information about $c_i$ and $x_i$. A detailed description of this measurement scheme is provided in Supplementary Material 1. 


To measure a violation of VBC's inequality, we now measure the individual terms of Eq. \ref{eq::LCI}.
For the first two terms, we only need to consider cases where $y=0$, which corresponds to Bob performing measurements in the computational basis. 
In the first term, we consider outcomes where he obtains $b=0$, which means that the photon is horizontally polarized, while in the second term Bob finds his photon to be vertically polarized which represents $b=1$.
We therefore set Bob to measure in both of these settings and for each we iterate over all combinations of the Alices' settings. We effectively trace over Charlie by having him perform measurements in the $Z+X$ and $Z-X$ bases and summing the results. For each possible combination of settings, we record the coincidence counts between Charlie’s four detectors and Bob’s detector in the transmitted output of his PBS.
Note, we discard the reflected outcomes since we only trigger our UFORs from the transmitted detector.
Since $x_1, x_2$ and $z$ take two possible values $\{0,1\}$, each of the first two terms is constructed from eight probabilities formed by the possible combinations of these settings. We then compute individual probabilities as 
\begin{equation}  
    p_1= \frac{1}{8} \sum_{x_1, x_2, z \in \{0,1\}} p(b=0, a_2 = x_1 |x_1 x_2 z, y=0 )
        \label{eq::term1}
\end{equation}
and \cite{van2023device}
\begin{equation}  
    p_2= \frac{1}{8} \sum_{x_1, x_2, z \in \{0,1\}} p(b=1, a_1 = x_2 |x_1 x_2 z, y=0 ).
        \label{eq::term2}
\end{equation}
For the last term in the inequality, Alice 1 and Alice 2 reprepare the target qubit in $\ket{0}$, which means their polarizers are set to transmit $\ket{H}$ and the waveplate after the polarizers are set to $0^\circ$. This ensures that the target qubit exits each of their setups with the same polarization $\ket{H}_{2_T}$ it had upon entering. We estimate Bob and Charlie's ability to win the CHSH game by iterating over their measurements constructing their winning probability as
\begin{equation}
    p_3 = \frac{1}{4} \sum_{y,z \in \{0,1 \}}p(b\otimes c = yz | yz).
    \label{eq::term3}
\end{equation}
Again, the normalization factor arises from the number of possible combinations of measurement settings.
For each measurement setting we measure the coincidences for 3 minutes, resulting in an average total counts of $\approx 7000$ per setting.
The individual probabilities are plotted in Fig. \ref{fig::Data}b), wherein the close agreement between the ideal values and our experimentally measured probabilities is apparent.

Summing these probabilities together, we observe a clear violation of VBC's inequality of $1.8328  \pm  0.0045$, which is 18 standard deviations above the Definite Causal Order Bound of $1.75$.
We attribute the small deviation between our measured violation and the theoretical quantum maximum of $\frac{6 + \sqrt{2}}{4} \approx 1.8536$ to imperfections in the entangled state and a slightly reduced process fidelity of the quantum switch.
To confirm this, {we model a reduced process fidelity of the quantum switch by introducing ``causally separable'' noise to the ideal quantum switch process matrix $W_{\text{switch}}$. In other words, we create the noisy  process matrix $W = (1-\epsilon) W_{\text{switch}} + \epsilon\left[W^{\text{A$_1$} < \text{A$_2$}} + W^{\text{A$_2$} < \text{A$_1$}}\right]$, and vary $\epsilon$ to fit our data.
When also including reduced purity of our input state, measured to be $ 0.98036 \pm 0.00034$ by performing quantum state tomography (Fig. \ref{fig::Data}a), we find that a process fidelity of $\mathrm{F_{Switch}}=0.9816\pm0.0069$ almost perfectly describes our measurement result (see inset of Fig. \ref{fig::Data}d), highlighting our high-fidelity implementation of the switch.
}

To provide additional insight into the VBC-inequality we next study how different types of noise affect it's violation.
To induce this noise in a controllable manner, we prepared each of the four Bell states and then carried out the original measurements with each of the Bell states as the input state. 
We then summed these different data sets together with different weights to mimic different noise.
We first studied the depolarizing channel, which can be written as 
\begin{equation}
\varepsilon \rho_{\Phi^+} = (1 - \eta)\ket{\Phi^+}\bra{\Phi^+}  + \frac{\eta}{4}\mathbb{I}, 
\label{eq::depol}
\end{equation}
where $\mathbb{I}$ is the identity matrix \cite{Nielsen.2010}. 
This destroys all correlations between Charlie and Bob as well as any coherence inside the Quantum Switch.

\begin{figure}[t]
    \centering
    \includegraphics[width=\linewidth]{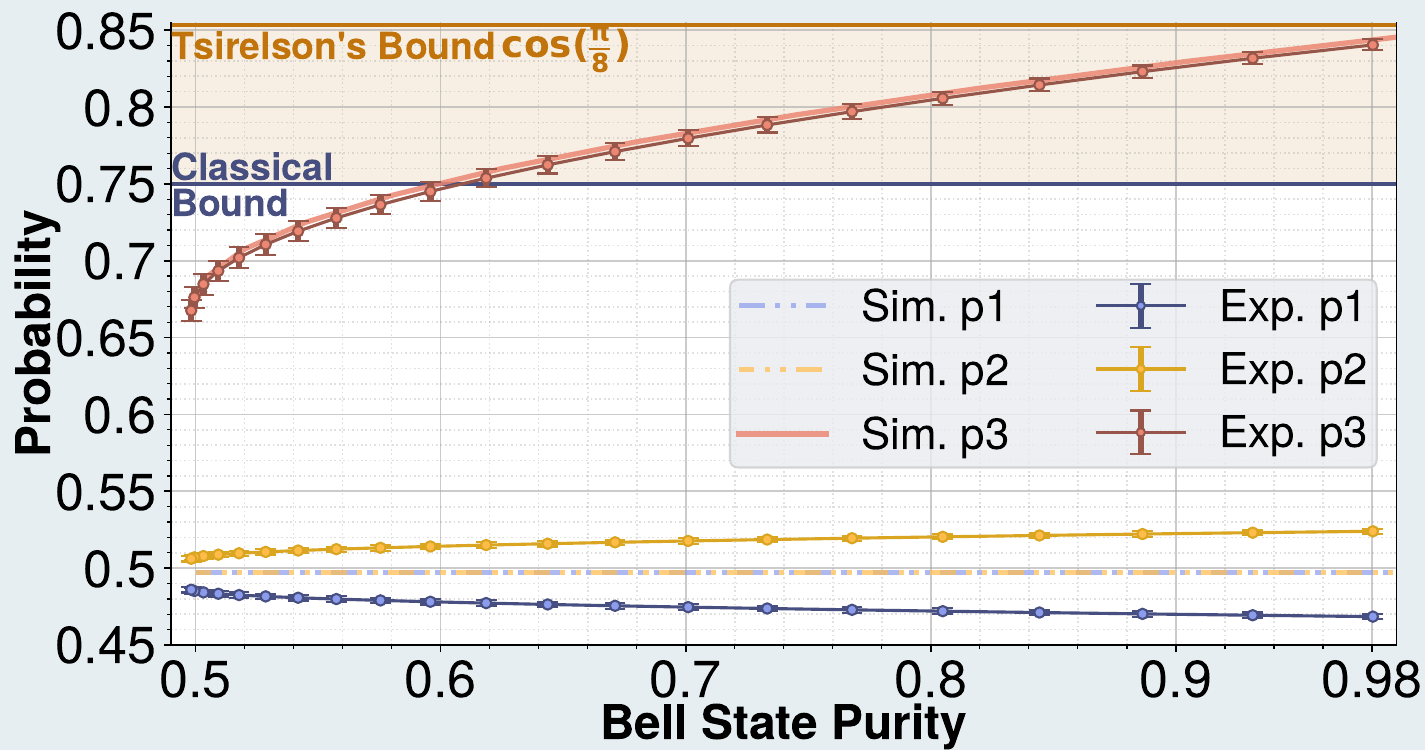}
    \caption{\textbf{ The effect of dephasing noise on the probabilities $p_1$, $p_2$ and $p_3$ of the VBC-inequality } Experimental data modelled with eq. \ref{eq::depahse} and simulated curves with a Switch process fidelity of $0.9816\pm0.0069$. }
     
    \label{fig::Dephasing}
\end{figure} 
Our results for applying depolarising noise, pictured in Fig. \ref{fig::Data} panels c and d, shows, that the ability to violate the VBC-inequality is strongly influenced by noise, reducing all three terms of the inequality (Fig. \ref{fig::Data}c). 
Since increasing depolarising noise destroys the entanglement between Bob and Charlie, it comes as no surprise, that the third term $p_3$ (Eq.\ref{eq::term3}), which represents a CHSH-Game (red line in Fig \ref{fig::Data} c)), falls bellow theDefinite Causal Order Bound of $0.75$. That it also reduces the probabilities $p_1$  and $p_2$ can be understood from the fact that these terms represent correlations between the ability of Alice 1 to signal to Alice 2 ($p_1$ Eq.\ref{eq::term1}) and vice versa ($p_2$ Eq.\ref{eq::term2}) to Bobs outcomes in the computational basis. The depolarizing noise not only removes the quantum coherence, but also these classical correlations.

The simulation of our experiment with  $\mathrm{F_{Switch}}=0.9816\pm0.0069$ agrees well our experimentally obtained VBC-value. The same holds true for the value of $p_3^{exp} = 0.8404 \pm  0.0036$. However, there is a discrepancy for $p_1$ and $p_2$, which deviate in opposite directions: $p_2^{exp}$ rises slightly above the ideal value and  $p_1^{exp}$ below it. This deviation is caused by imperfections in the Bell state generated by our photon source. Although the state has a high purity, it has a slight imbalance in the  $\ket{VV}$ and $\ket{HH}$ terms, which results in an asymmetry $p_1$ and $p_2$ since Alice 2 and signals to Alice 1 slightly more often.

We also studying dephasing noise defined as 
\begin{equation}
    \zeta \rho_{\Phi^+} = (1 - \vartheta)\ket{\Phi^+}\bra{\Phi^+}  +  \vartheta\ket{\Phi^-}\bra{\Phi^-}
\label{eq::depahse}
\end{equation}
with a phase flip-probability $\vartheta$ \cite{Nielsen.2010}\cite{Wilde.2017}. Where $\vartheta = 0.5$ corresponds to the maximum phase noise.
The VBC inequality behaves differently in this case. Again it reduces the purity of the joint state and thereby destroys the entanglement and decreasing $p_3$, the winning probability of the CHSH game (red line in Fig.\ref{fig::Dephasing} a)). However, since it only removes coherence between Charlie and Bob, they maintain classical correlations. Therefore, the correlations between Bobs’ results and the causal order inside the quantum switch remains intact. As a result $p_1^{sim}$ and $p_2^{sim}$ remain constant for any value of $\vartheta$ (blue and yellow lines in Fig.~\ref{fig::Dephasing} b)). 
The experiment agrees well with our simulations, with the slight deviations in $p_1$ and $p_2$ again attributed to imperfections in the generated Bell states.

\section{Discussion}
While the protocol that we have implemented is device-independent, our experimental implementation contains loopholes that must be closed to achieve a fully device-independent verification of indefinite causal order.
Since Bob and Charlie's CHSH game lies heart of VBC's inequality, the standard Bell loopholes \cite{Shalm2015strong,Giustina2015Significant,Hensen2015Loophole} must be closed. These include  measurement independence (freedom of choice), fair-sample (detection), and the locality loophole. Since our experiment is a proof of principle, we did not close these loopholes. Nonetheless, we will briefly mention how they manifest in our setup. To ensure measurement independence, the measurement settings must be either random or freely chosen \cite{Scarani.2019,Bell_Aspect_2004}. Our implementation does not satisfy this requirement, as the measurement order was predefined in our code. The fair-sampling loophole arises when a potentially unrepresentative subset of events is detected, which may falsely suggest a Bell violation. In the present implementation, our setup experiences significant loss, mainly due to multiple passes through the UFORs and the the relatively low transmission of the polarizers used for the Alices' measurements. In total, the net detection efficiency through the entire experiment is $\approx 1\%$, far from closing this loophole. Closing the locality loophole requires space-like separation between Bob and Charlies measurement events \cite{Scarani.2019}. 
Our experiment is built on a single optical table with measurement stages less than 1m apart. 
Moreover, Bob's photon is detected first, and used as a timing reference for the UFORs.
To close this loophole, we would need to work with a pulsed source of entangled photons and synchronize the UFORs to the source. This would allow us to separate the other parties from Bob by a much greater distance and perform the measurements without any time delay.
Using this timing method would further strengthen the experiment by allowing the measurement order to match the theoretical VBC-scenario, as registering Bob’s outcome earlier introduces a potential causal link from Bob to the other parties.

While we know how to close the standard Bell loopholes, the verification of ICO opens new loopholes.
Here we point out two loopholes related the definition of time-delocalised events \cite{oreshkov2019timeDelocalized, vilasini2025events}.
The first is  specifically related to VBC's inequality.  As we sketched in Fig.\ref{fig::idea} a) Alice 1 and Alice 2 must act in the past light cone of Charlie.
While the photon certainly passes through the Alices' labs before reaching Charlie, in our implementation (and in all implementations of measurements in a switch \cite{rubino2017ExperimentalVerification,cao2022Semideviceindependent}) the measurement results are not read out `locally' in their labs.
However, this is, in principle, possible to achieve in a purely quantum optical setting.
For example, one could implement a quantum non-demolition measurement of the photon's polarization by coupling the photon to an auxiliary probe system that is stored locally in each lab using a single-photon level nonlinearity \cite{nogues1999seeing}.
If this is not done properly it will introduce which path information, decohering the switch. 
However, by coupling the probe system to each mode in the local labs one can realize read out the information locally without yielding which-path information.
This could be done similar to proposed gedanken experiments that count the gate uses in the quantum switch \cite{Procopio2015Experimental,rozema2024experimental}.

A related but distinct loophole is enforcing the closed lab assumption \cite{oreshkov2012quantum}.
In the closed lab assumption, each party is imagined to act in an isolated lab with an input door and output door.
Each door is opened exactly once to ensure the party acts once.
Since our photon's coherence time is shorter than the time required for the photon to propagate between Alice 1 and Alice 2's labs there are two distinct times at which each party might act.
This follows from the fact that the photonic switch is often said to have 4 distinct space-time events\cite{goswami2018Indefinite,rozema2024experimental}.
Therefore, it would be possible for Alice 1 to open and close her lab to let the photon in at time 1 and then again at a second time.
This would not have any observable effect on the experiment, which could allow us to conclude the photon entered the lab twice.
The situation is different when the photon is temporally delocalized.
For example, if the photon has a coherence time longer than the propagation time between the labs, then doors could be only opened once without distrubing the photon and changing the outcomes.
One experiment \cite{goswami2018Indefinite} has realized this, albeit with unitary operations in the switch.
Although using temporally long photons may be a step in the right direction for the closed lab assumption, enforcing this in a loophole-free experiment is another matter, that we leave for future discussion.

We have used a time-bin implementation of the quantum switch to violate a device-independent inequality, indicating the presence of ICO between two parties in our experiment {and showed how it responds to different types of experimentally relevant noise.}
Our strong violation of $1.8328  \pm  0.0045$, close to the theoretical maximum bound of $1.8563$ is made possible by our passively-stable high-fidelity experiment, showing that it is a promising implementation of the quantum switch both for foundational tasks such as that discussed here and to implement ICO-based advantages.
This represents an important step towards a loophole-free verification of ICO,
which is crucial for supporting photonic quantum switch experiments, as there currently is not a consensus as to whether such experiments realize a genuine ICO or merely simulate it. For instance, some argue that a genuine ICO can arise only from the so-called gravitational quantum switch \cite{zych2019bell}, asserting that ICO requires superpositions of gravitational fields \cite{paunkovic2020distinguishing}. The other side, in contrast, supports existing experiments, maintaining that the ICO in photonic quantum switch experiments originates from delocalized events which is equivalent to a superposition of spacetimes that depends solely on the choice of quantum coordinates \cite{DeLaHemette2022quantumDiffeomorphisms}. Closing the loopholes in the violation of the VBC inequality presented here would finally settle this debate.
Furthermore, it would confirm that ICO is a new quantum resource distinct from entanglement and would provide a footing for the many recently-proposed protocols exploiting ICO to accomplish tasks that cannot be carried out with standard quantum processes. 

\nocite{supplemental}
\nocite{Zanin2022enhancedphotonic}\nocite{LuizZanin:21}

\section{Acknowledgements} 
This project has received funding from the European Union (ERC, GRAVITES, No 101071779), the European Union’s Horizon 2020 research and innovation programme under grant agreement No 899368 (EPIQUS), the European Union’s Horizon 2020 research and innovation programme under the Marie Skłodowska-Curie grant agreement No 956071 (AppQInfo) and the European Union (HORIZON Europe Research and Innovation Programme, EPIQUE, No 101135288). Views and opinions expressed are however those of the author(s) only and do not necessarily reflect those of the European Union or the European Research Council Executive Agency.
This research was funded in whole or in part by the Austrian Science Fund (FWF)[10.55776/COE1] (Quantum Science Austria), [10.55776/F71] (BeyondC) and [10.55776/FG5] (Research Group 5). For open access purposes, the author has applied a CC BY public copyright license to any author accepted manuscript version arising from this submission.
This material is based upon work supported by the Air Force Office of Scientific Research under award number FA9550-21-1-0355 (Q-Trust) and FA8655-23-1-7063 (TIQI).
The financial support by the Austrian Federal Ministry of Labour and Economy, the National Foundation for Research, Technology and Development and the Christian Doppler Research Association is gratefully acknowledged.
\\

\section{Data and Code availability} {All the data and code  that are necessary to replicate, verify, falsify and/or reuse this research is available online at \cite{richter_2025_15704777}.}\\

\bibliography{main}

\clearpage

\section{Supplementary Material}

\subsection{Time-Bin Quantum Switch}
\label{sec:switch}

\begin{figure*}[ht!]
    \centering
    \includegraphics[width=0.8\textwidth]{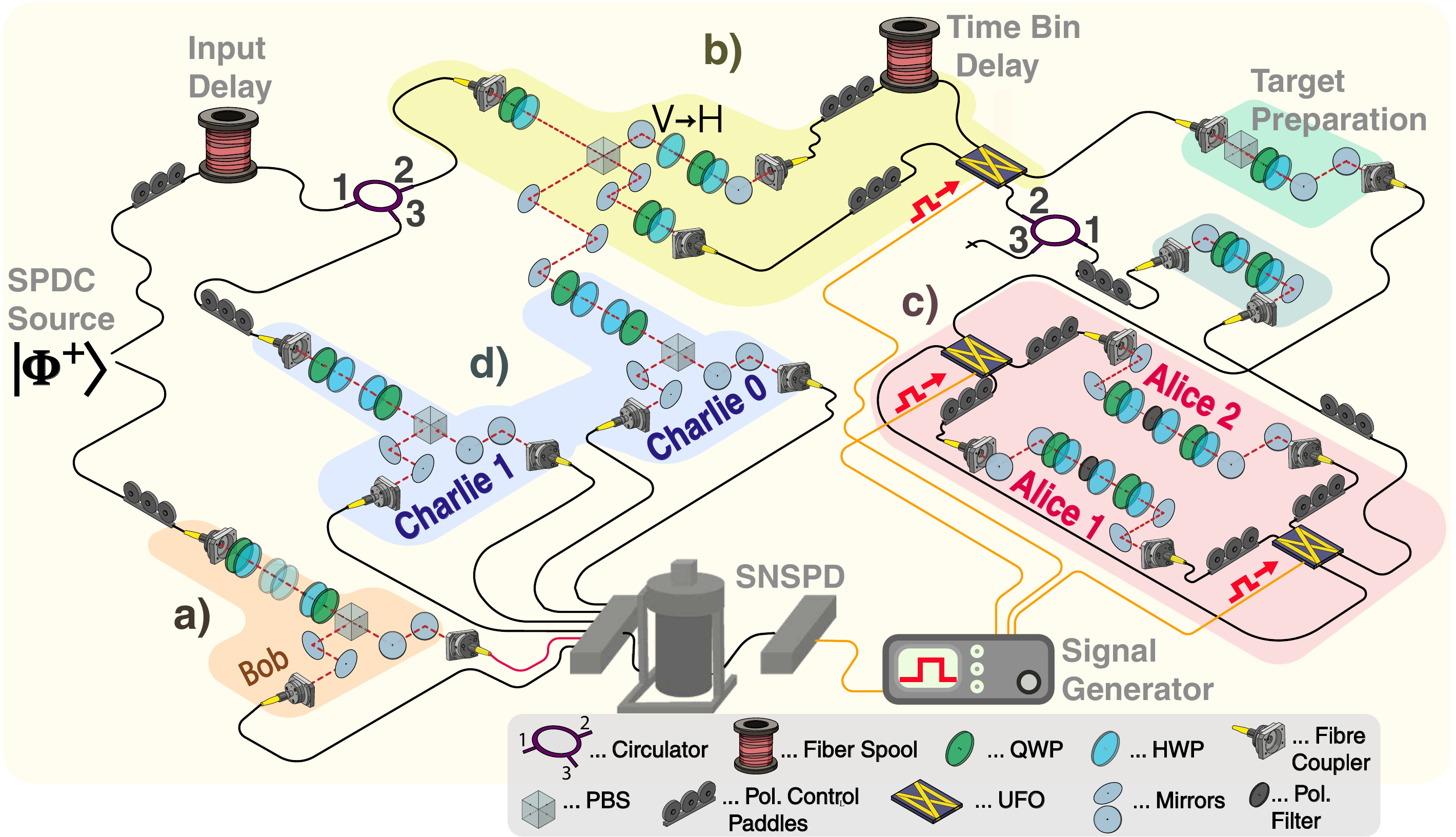}
    \caption{\textbf{Schematic image of the full experimental setup:}
\textbf{a)} Bob’s tomography stage consisting of a PBS and a set of waveplates (QWP + HWP). Optionally, one or two additional HWPs can be inserted in front of this measurement setup to transform the Bell state $\ket{\Phi^+}$ into any other Bell state.
\textbf{b)} Interferometer with unequal arm lengths: on the forward path it converts the photon’s polarization qubit into a time-bin qubit, and on the return path it allows simultaneous measurement of target and control qubits.
\textbf{c)} Fiber-based quantum switch.
\textbf{d)} Charlie’s measurement setup, split into two tomography stages.
\textbf{Note:} Each fiber section is polarization-compensated using a set of waveplates and polarization controller paddles. }
    \label{fig:FullSetup}
\end{figure*}

In this work, we use the passively stable fiber-based quantum switch first presented by Antesberger et al. in Ref. [22]. What sets this architecture apart from other implementations of the quantum switch is that it employs time-bin qubits as the control instead of spatial modes, so that only a single spatial mode needs to traverse each optical element, significantly simplifying the setup. Passive stability is ensured by using a single interferometer for both preparation and measurement of the control system, thereby eliminating all phase fluctuations by design. 
This allows measurements over long periods without the need of adjustments, enabling measurements to be carried out over long acquisition times.

To generate the time-bin qubit that controls the order of operations, we start by creating a pair of photons at a wavelength of 1550 nm using spontaneous parametric down-conversion (SPDC). One photon is send to a measurement stage (called Bob) where it's polarisation state is measured. At the same time, this measurement provides a time reference for the rest of the experiment. 

After the first photon is immediately detected by Bob, the second photon is sent to a polarizing beam splitter (PBS), which separates into two paths based on its vertical and a horizontal polarization components. The vertically polarized component is sent through a half-wave plate (HWP) that rotates its polarization to horizontal and then through a fiber spool that introduces a 150 ns delay relative to the other component, which takes a shorter path.

We then recombine these two paths deterministically using an ultrafast fiber-optic switch (UFOS), as shown in Fig \ref{fig:FullSetup} and \ref{fig:meas_fig} (lime green shaded areas).  An electronic pulse, triggered by the detection of the first photon, controls the UFOS in the time-bin interferometer and the quantum awitch. Triggered by this detection event, the UFOS operates such that both time-bin qubits reach the same interferometer, which first routes the short-path component and then the delayed long-path component into the same output mode of a Mach–Zehnder interferometer. This results in the second photon being in a coherent superposition of two time bins within a single fiber.

We use BATi 2×2 Nanona fiber switches as UFOS devices, which serve both to generate the time-bin qubit and to control the photon routing within the quantum switch. The switches exhibit a 60 ns response time, support switching rates up to 1 MHz, and provide isolation exceeding 20 dB between output channels for any polarization (see Refs. [56], [57]]).

After preparing the time-bin qubit, we implement the quantum switch operation on the target system, encoded in the polarization of the same photon. This is achieved by applying a pulse sequence of three low and two high voltage levels to the UFOSs. In the low-voltage ``cross'' state, input ports are directed to opposite outputs, while in the high-voltage ``bar'' state, inputs pass straight through. The fiber lengths in the quantum switch are precisely adjusted to ensure correct timing and synchronization. For a detailed description of this method, we refer to the aforementioned publication [22].

Controlling the routing in this manner causes the photon’s polarization--the target system--to undergo Alice 1’s and Alice 2’s operations in a superposition of different orders, determined by the state of the time-bin control qubit, which is entangled the polarization of the first photon detected by Bob. 

After the quantum switch, the time bins pass through the interferometer again in the opposite direction, where they are recombined at the PBS and subsequently the photon is measured in Charlie’s tomography stages (see next section).

\newpage

\subsection{Control Qubit Measurement}

\begin{figure}[b]
    \centering
    \includegraphics[width=\linewidth]{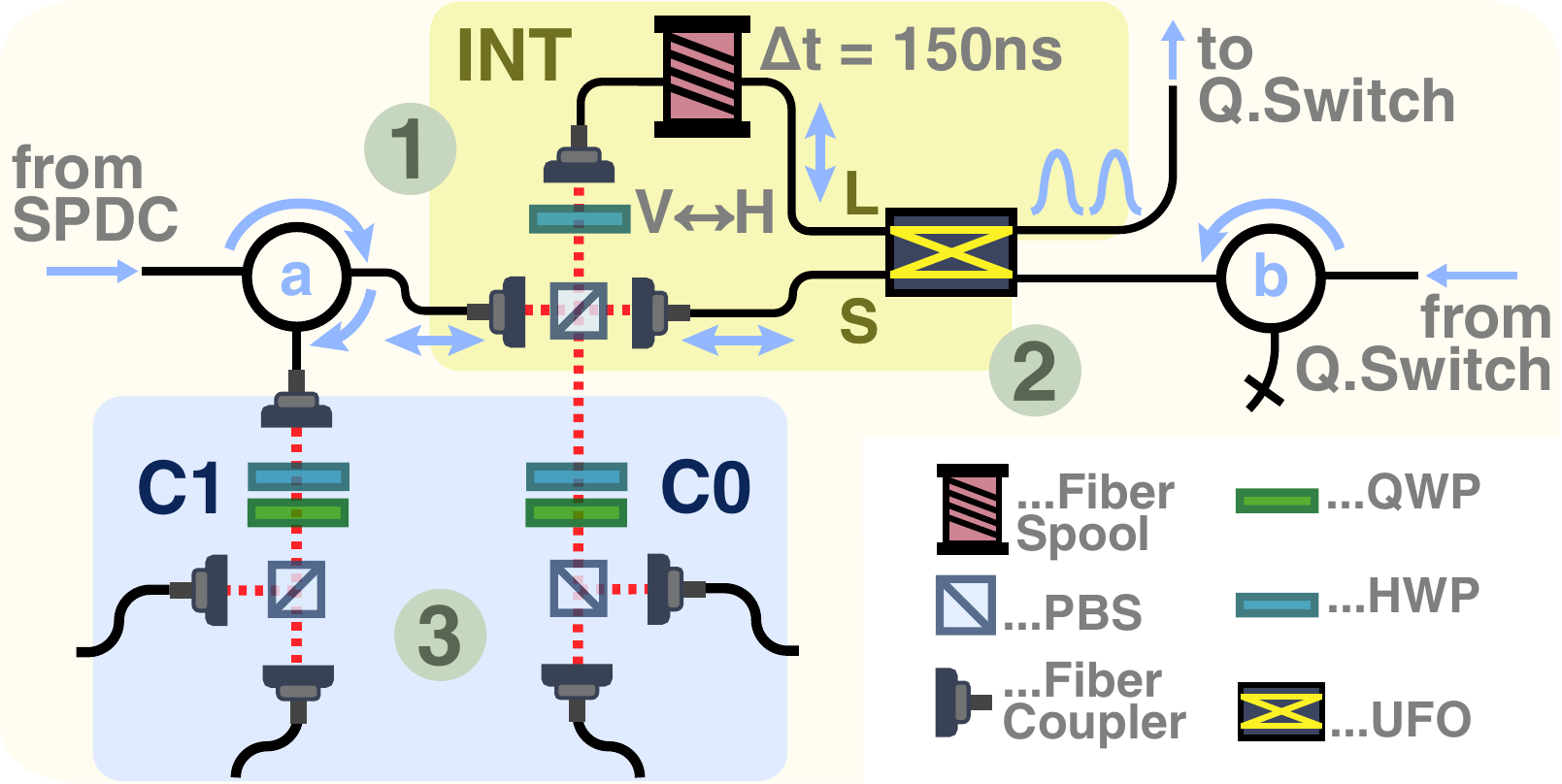}
    \caption{\textbf{Schematic image of the time-bin interferometer and Charlies measurement stage.} \textbf{1:} The circulator (a) directs the photon into the interferometer (INT), where its polarization degree of freedom is transformed into a time-bin qubit with horizonatl polarisation. \textbf{2:} After the quantum switch, the photon is directed back into the interferometer through the second circulator (b). The deterministic UFO switches for the time bins such that they recombine at the PBS in the interferometer. \textbf{3:} Depending on its previous path and it's polarization, the photon is detected in one of Charlie's four detectors.}
    \label{fig:meas_fig}
\end{figure}

Our setup allows us to measure the control and target qubits simultaneously by splitting Charlie’s measurement setup into two tomography stages. This is made possible by the following method: When the second photon travels from the SPDC source towards the interferometer section (Fig.~\ref{fig:meas_fig} (1)), it passes a circulator (a) that directs it to the interferometer (yellow area in Fig. \ref{fig:meas_fig}), where it arrives e.g., in the polarization state $\ket{\phi_{in}}=\tfrac{1}{\sqrt{2}}\big(\ket{H}+\ket{V}\big)$. From here, it reaches a PBS that separates its polarization components into two paths with a path-length difference of 150 ns. The vertically polarized component is sent into the longer path (L), where a HWP rotates its polarization to horizontal. The short (S) and long (L) paths are then recombined by an ultrafast optical switch (UFOS), yielding two qubits encoded in a single photon, in its time and polarization degrees of freedom. The time-bin qubit represents the control in a superposition of early ($\ket{E}_C$) and late ($\ket{L}_C$) arrival time, the now uniformly horizontal polarization of the photon serves as target system ($\ket{H}_T$) for the operations inside the Quantum Switch : $\ket{\phi'_{in}}=\frac{1}{\sqrt{2}}\big(\ket{E}+\ket{L}\big)_C\ket{H}_T$.
Depending on their arrival time the photon experiences these operations performed by Alice 1 (A1) and Alice 2 (A2) in 
two different orders, that result in different polarisation states of the target qubit after the Quantum Switch, therefore we receive the output state $\ket{\phi_{out}}=\frac{1}{\sqrt{2}}\big(\ket{E}_C\ket{\psi_{(A_1\rightarrow A_2)}}_T+\ket{L}_C\ket{\psi_{(A_2\rightarrow A_1})}_T\big)$.
The photon returns to the interferometer and reaches the UFOS (Fig.~\ref{fig:meas_fig} (2)). Here the switch is timed to the photon’s arrival, ensuring that the time-bin qubits are routed back to the PBS with the early bin taking the long path and the late bin taking the short path. Thereby the photon components are reunited in time, when they arrive at the PBS, with the control information encoded in the path degree of freedom ($\ket{E}_C \rightarrow \ket{P_L}_C$,  $\ket{L}_C \rightarrow \ket{P_S}_C$): $\ket{\phi'_{out}}=\frac{1}{\sqrt{2}}\big(\ket{P_L}_C\ket{\psi_{(A_1\rightarrow A_2)}}_T+\ket{P_S}_C\ket{\psi_{(A_2\rightarrow A_1})}_T\big)$. Depending on the path, the polarization components are now split into one of two paths leading to Charlie’s two tomography stages, C0 and C1 (blue area in Fig.~\ref{fig:meas_fig}, (3)).
If the photon arrives from the short path, it carries the information of the former late time-bin qubit, which arrived first at Alice 2 and then at Alice 1 inside the Quantum Switch. Its vertically polarized part now reaches tomography stage C0, while the horizontal part travels back through the circulator (a) to tomography stage C1.
For the former early time-bin qubit, the horizontal polarization arrives at C0, while the vertical polarization is routed to C1.
\label{sec:measurement}
\subsection{Quantum State Tomography of the Resource state}
\label{sec:tomography}
\begin{figure}[th]
    \centering
    \includegraphics[width=\linewidth]{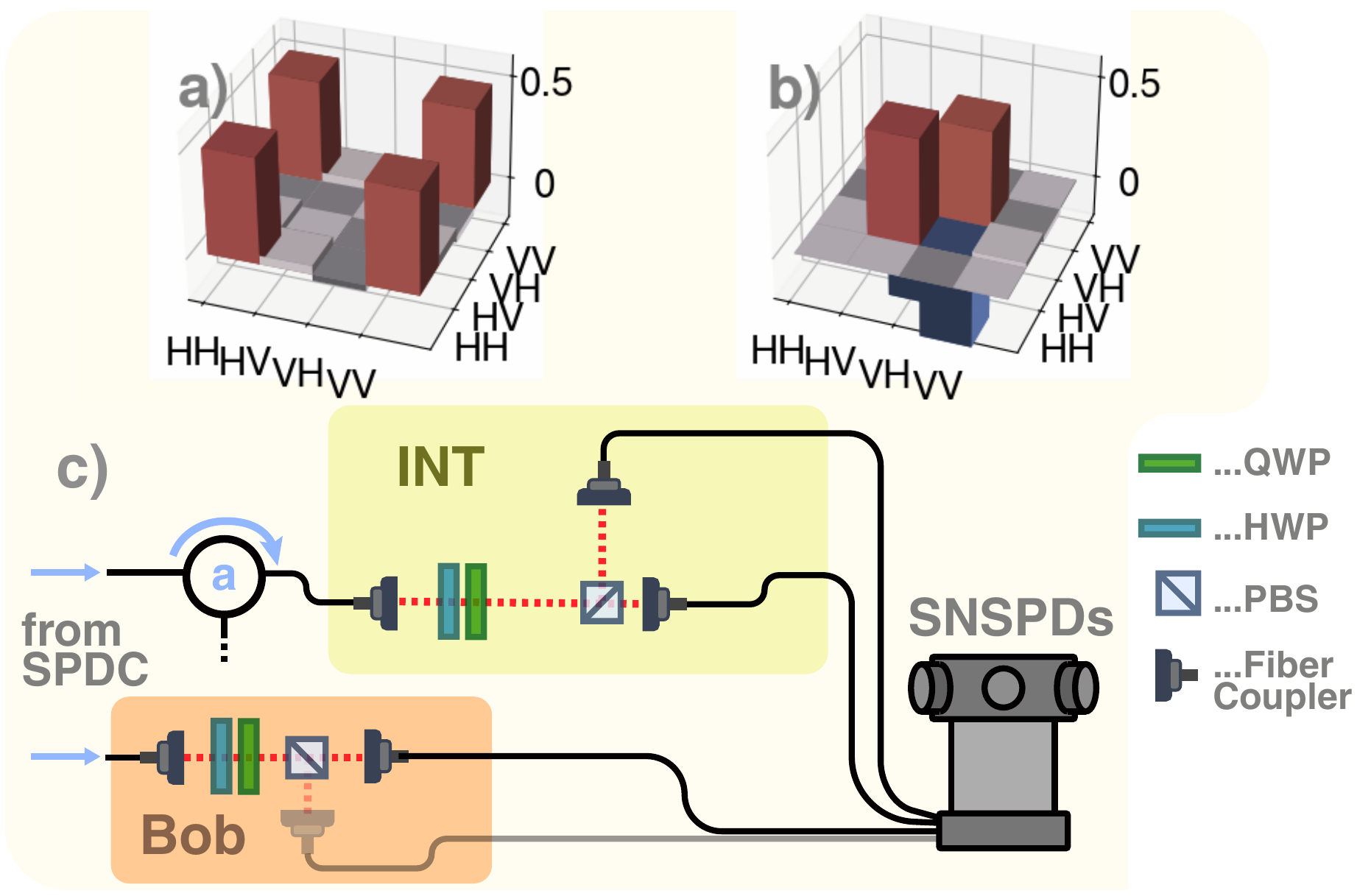}
    \caption{\textbf{Density Matrix of the two-photon state before and after the second photon has passed the Quantum Switch} \textbf{a)} The input state with a purity of $0.98036 \pm 0.00034$ and a concurrence of $0.9792\pm 0.0011$.  \textbf{b)}The two photon state after the quantum Switch with a purity of $ 0.9388 \pm 0.0040$ and a concurrence of $ 0.9352 \pm 0.0042$. \textbf{c)} Schematic sketch of the modified setup for the two-qubit tomography measurement on the resource state before it enters the quantum switch.}
    \label{fig:rhos}\end{figure}
We performed two-qubit quantum state tomography to assess the quality of the two-photon state generated by our SPDC source and to verify the functionality of our setup. The tomography was conducted between Bob and the second photon—once before the photon entered the quantum switch, and again after it reached Charlie’s measurement stage. To enable the tomography on the state right after the source, we adapted the interferometer to act as a tomography stage. This was achieved by inserting a quarter-wave plate (QWP) and a half-wave plate (HWP) directly before the polarizing beam splitter (PBS), and by removing the HWP from the reflected path. The output couplers of both interferometer arms were then connected via optical fibers directly to the superconducting nanowire detectors (SNSPDs). Since we only use the transmitted counts from Bob’s PBS for the VBC-inequality measurement, we also only used those counts for the tomography. Therefore, we projected the eigenstates of all bases onto horizontal polarization.
Charlie’s setup already includes two tomography stages, so no changes were needed after the quantum switch. For each measurement setting One detector from Charlie 0 and one detector from Charlie 1 must be counted together, representing a single detection event. During this measurement, the HWPs and polarization filters of Alice 1 and 2 inside the quantum switch are rotated to their zero positions, while the UFOs switch in the same synchronization as in the VBC-inequality measurements.

\end{document}